\documentclass[12pt,draftclsnofoot,onecolumn]{IEEEtran}
\usepackage{graphicx} 

\usepackage{amsmath}
\usepackage{enumerate}
\usepackage{amsthm}
\usepackage{amsfonts}
\usepackage{setspace}		
\usepackage{longtable}          
\usepackage{multirow}
\usepackage[usenames,dvipsnames]{color}
\usepackage{amsmath}
\usepackage{mathtools}
\usepackage{bm}
\usepackage{graphicx}
\usepackage{caption}
\usepackage{subcaption}
\usepackage{stfloats}
\newcommand{\rev}{\color{black}}

\begin{document}

\title{EEG Compression of Scalp Recordings based on Dipole Fitting}

\author{H. Daou,~\IEEEmembership{Student Member,}
        F. Labeau,~\IEEEmembership{ Senior Member, IEEE}
\thanks{This work was supported by the Natural Sciences and Engineering Research Council (NSERC) and industrial and government partners, through the Healthcare Support through Information Technology Enhancements (hSITE) Strategic Research Network.}
\thanks{This work has been submitted to the IEEE for possible publication. Copyright may be transferred without notice, after which this version may no longer be accessible}}

\maketitle

\begin{abstract}
A novel technique for Electroencephalogram (EEG) compression is proposed in this article. This technique models the intrinsic dependency inherent between the different EEG channels. It is based on dipole fitting that is usually used in order to find a solution to the classic problems in EEG analysis: inverse and forward problems. The suggested compression system uses dipole fitting as a first building block to provide an approximation of the recorded signals. Then, (based on a smoothness factor,) appropriate coding techniques are suggested to compress the residuals of the fitting process. Results show that this technique works well for different types of recordings and is even able to provide near-lossless compression for event-related potentials.  
\end{abstract}


\IEEEpeerreviewmaketitle

\section{Introduction}

Electroencephalography is the monitoring or recording of electrical activity in the brain. Electroencephalogram (EEG) recording can result in huge amounts of data to be stored and/or transmitted, which calls for efficient compression techniques \cite{SiraamNN} \cite{Dauwels}.

EEGs are used to visualize and analyze brain activity. While reading EEG signals, redundancy is highly visible in a single channel between different time segments, and between different channels. This correlation should be exploited when building a compression algorithm. In the following paragraphs, different methods that have been recently suggested to compress EEG signals are presented.

Predictors, such as linear predictors (LPs) \cite{lp_1994}, and neural networks predictors both using lossless \cite{Sriraam2012379} \cite{Sriraam:2012}, and near lossless \cite{SiraamNN} algorithms, have been suggested to compress EEG signals. In addition, a combination of both neural network and linear predictors is also used to achieve lossless compression of EEG signals \cite{sriraam_eswaran_2006,sriraam_eswaran_2008}. In these models, a predictor is used in the first stage to decorrelate the source data by removing the correlation between neighbouring samples.

Iterative function sets and genetic algorithms are also used to compress EEG signals \cite{mitra_1997}. This compression technique involves the use of sets of linear transformations to provide an approximation of the signal. Using genetic algorithms (GAs), self-similarities in the EEG signals are explored to identify the proper sets of transformations. Gain/Shape vector quantization is also used to approximate the EEG signals \cite{gurkan_2009}. This technique uses classified signature and envelope vector sets and assumes both transmitter and receiver share the same sets.

Many compression algorithms use Wavelet Transform (WT) to decompose a signal and take advantage of the properties of these coefficients in energy compaction. WT provides multi-resolution, locality, and compression when combined with zero-tree coding techniques. It was shown that choosing an appropriate mother wavelet gives better performance results than when using an arbitrary wavelet \cite{optwav}, \cite{srinivasan:25}.

Another type of WT where the signal is passed through more filters, Wavelet Packet Transform (WPT), is used by Cardenas-Barrera et al. to segment and decompose the EEG Signals \cite{valdivia_2004}. The compression algorithm is mainly composed of thresholding of the low-relevance coefficients, then applying quantization and Run-Length Coding (RLC). However, calculating the proper threshold is the main issue in this model. 

WT can be combined with efficient coding techniques, such as embedded zero-tree wavelet (EZW) \cite{Dehkordi}, that take advantages of this transform's characteristics. Set Partitioning in Hierarchical Trees (SPIHT) has also been used, both in $1$D \cite{elec_let} \cite{journal}, and $2$D \cite{Hoda1} \cite{srinivasan:26}, to compress scalp EEG recordings. In fact $2$D-based methods explore both time and spatial dimensions when coding the signals, and thus are able to capture more redundancy and similarities.

$3$D Transforms are also used to further decorrelate the signals and provide efficient compression \cite{Dauwels}. To perform compression, the authors consider wavelet-based volumetric coding, energy-based lossless coding and tensor decomposition based coding. Afterwards, the residual signals are coded either in a lossy \cite{Dauwels} or lossless \cite{Dauwels2} fashion then transmitted.

Overall, these schemes are based on applying different transforms and coding schemes in order to extract the inherent redundancy present in both the spatial, i.e. between different channels, and the temporal, i.e. between different time samples, domains. In this article, we suggest exploring the inverse problem of EEG recordings and applying it in the context of compression. In other terms, we propose modelling the recorded signals using the source dipole that generates these waveforms on the scalp of the head (see section II for details).
Therefore, by using a physiologically meaningful model, we hope to get a better space for EEG signal representation in order to achieve better approximation for a low bit rate. Our experiments show the superior performance of this method compared to methods based on classic transforms.

In this article, a method to localize the sources behind the recorded electrical activity on the scalp is explored. First, the forward/inverse problems are briefly introduced. Then a technique used to localize the source dipoles and build the forward model that maps these dipoles to the observed scalp recordings is described in details. At the end, compression methods of the dipole fitting residuals and moments are suggested and results are shown for three different databases of recordings, having different characteristics.

\section{EEG Inverse and Forward Problems}
The non-invasive localization of the neuronal generators that are behind the observed EEG signals, which is known as inverse solution, is one of the main concerns in electrophysiology \cite{review}. Relying on the pattern of EEG recorded at the scalp’s surface to determine the generators is a big challenge and of great interest \cite[p.73]{Current}. Finding a solution of the inverse problem is able to give us a model that maps the generators to the recorded projections.

Thus the inverse problem aims at finding the true functional tomography by localizing the sources with minimum error. However, the main challenge that such a problem faces is that the measurements do not contain enough information about the generators, which makes the problem ill-posed and thus a perfect tomography can not exist. The reason behind such challenge is that different internal source configurations can generate the same external electromagnetic fields and these fields are measured only at a relatively small number of locations \cite{review}.

The cerebral cortex, which is the outer surface of the brain, is comprised of 10 billion neurons. Most observed scalp activity is generated within this part of the brain which is $1.5$ to $4.5$ mm thick. A synchronous synaptic simulation of a very large number of neurons results in a dipolar current source oriented orthogonal to the cortical surface \cite{ica_zhukov}. The measured EEG is actually the propagation of this current onto the different electrodes' locations.

In order to compress EEG recordings, one aims at estimating the scalp recordings while minimizing a certain chosen distortion criterion. Having solved the inverse problem, one can use such model to generate, from the calculated dipoles, an approximation of the EEG recordings. This is known as the forward problem.

Therefore, we first analyse the EEG signal by using the inverse problem, then an approximation is generated using the forward model. This technique is somewhat close to the Analysis-by-synthesis techniques. 

 In order to define the EEG forward problem, one needs to well define the head geometry or model, the exact electrode positions within this geometry and the calculated current dipoles. Details of the method are presented in the following sections.

\section{Dipole Fitting}
To model the EEG recordings using the forward problem, the relationship between the distribution of the different primary current dipoles and the observed data at the sensors or electrodes needs to be defined \cite{forwp1}. There are many different head models that aim at approximating the volume conduction inside the head in order to find a solution to the inverse problem. In this section, descriptions of the electrode positions, head models, lead field and forward model algorithm are presented.

EEG inverse problem aims at finding and localizing the source behind the observed electrical activity which is important in the diagnosis and treatment of illnesses. Fur such cases, accuracy in building and solving the model is very important. However, in the context of compression, the solution to the inverse problem is used simply to compute approximations of the measured signals. Thus, high precision in dipole localization is not mandatory. For this reason, approximations about the electrodes' positions, head models, and other parameters that influence dipole localization can be tolerated.

It should be noted that the inverse and forward models are based on the dipole fitting algorithm of the Field Trip toolbox which is a Matlab toolbox for MEG and EEG analysis that includes algorithms for source reconstruction using dipoles, distributed sources and beamformers \cite{ft1}. The algorithm is modified and adapted to fit into the context of compression, details can be found in section III-C.

\subsection{Electrode Positions}
EEG electrodes are usually located on the scalp according to the International $10$-$20$ System of Electrode Placement \cite[p.139]{EEGB}. Spacing between electrodes is standardized in function of the horizontal and vertical widths of the head of the patient. Thus approximations of the positions can be easily calculated by simply using the standardized nomenclature of the electrodes. However, certain EEG labs provide the specific electrode positions with the EEG recordings which makes the step of specifying the electrode positions on the scalp straightforward and more precise.

Exact electrode positions associated to different channels can be easily found published online with specific recordings. These positions can either be in the polar, linear or spherical coordinates.  Mapping is done according to the name of the channels that is based on the International $10$-$20$ System. These positions can be directly used when referential montages are used in the recording.

As mentioned previously, high precision in dipole localization is not mandatory. For this reason, approximations about the electrodes' positions can be tolerated. Testing showed that using approximations for the electrodes' positions, even for bipolar montages, can still provide a rough dipole localization that gives good approximations of the signals. This will be further discussed in section VI.

\subsection{Head Models}
There are many different head models that aim at approximating the volume conduction inside the head in order to find a solution to the inverse problem. The simplest model is the single homogeneous sphere where the head is considered to be a homogeneous sphere of a certain radius $R$ and conductivity
$\sigma$. The potential at radius $r_d$ = $R$ for all $\theta$ (azimuth) and $\psi$  (latitude) in the coordinate system ($r_d$,$\theta$,$\psi$ ), that results from a dipole with moments ($G_x$, $G_y$, $G_z$), located along the z-axis at $r_d$ = $R$ can be easily calculated using equation~\eqref{eq:1} of finding the potential at the surface of a homogeneous sphere in \cite{robk}.

Homogeneous spheres neglect the effects of conductive inhomogeneities and irregular geometry that characterize the cranium \cite{robk}. 
These inhomogeneities are able to take into account the attenuation and smearing of the scalp potentials when performing dipole fitting. Thus, they can provide more accuracy in finding the dipoles' positions and amplitude.  

Inhomogeneous spheres are suggested to account for these irregularities. In this model, the head model is divided into different compartments that are limited by concentric spheres as follows:
\begin{enumerate}
\item The brain and the Cerebrospinal fluid (CSF) are modelled as a spherical volume of radius $r_{d_1}$ and
conductivity $\sigma_1$;
\item The skull is modelled with conductivity $\sigma_2$ and thickness $r_{d_1} - r_{d_2}$;
\item The scalp is modelled as a layer of conductivity $\sigma_3$ and outer radius $r_{d_3}$.

\end{enumerate}

Different numbers of compartments can also be used. The surface potential, at the different electrode locations is computed at the boundary $r_{d_3}$. Certain studies argue that, by comparing both models, the homogeneous assumption effect does not cause a degradation of the estimation problem \cite{robk}. However, the nested concentric sphere model is more commonly used since it better approximates the volume conductor model.

A third model assumes a more realistic head shape than a simple sphere by using information from anatomical images of the head. This model is found using a multiple compartment boundary element method (BEM) from magnetic resonance images (MRIs). This BEM model uses realistically shaped compartments of isotropic and homogeneous conductivities to better approximate the volume conductor properties compared to the simple spherical shells \cite{bem1}. The main problem is that the specific patient's MRI is needed in order to compute the model. However, research has shown that a standardized boundary element method (sBEM) volume conductor model that is derived from an average MRI dataset (from the Montreal Neurological Institute) can be used instead of patient specific models \cite{bem1}.

The standardized BEM model is a compromise between the simple spherical model that neglects the shape of the head, and the actual BEM model that requires a description of all brain compartments of the specific individual and requires a lot of computation and memory. In section IV, we compare the effect of the two head
models: concentric spheres with 4 compartments and standard BEM head model in order to choose the appropriate model for our algorithm.

To determine the appropriate head model, the DIPFIT validation study in \cite{eeg_online} is performed.

\subsection{Combined Inverse and Forward Models Algorithm}

This section describes the method used to estimate the EEG recordings using the inverse and forward models. 

In the remainder of this paper, when the recorded data is segmented into blocks along the time dimension, $\mathbf{s}_l[i,n]$ refers to the recorded sample at channel $l$, block index $i$ and index $n$ within block $i$. However, when no segmentation is applied, the recorded data is referred to as  $\mathbf{s}_l[n]$ where $n$ is the time sample computed since the start of recording (i.e. $\mathbf{s}_l[i,n]$ = $\mathbf{s}_l[iN+n]$ with $N$ equal to the block size). In addition, it should be noted that in this paper symbols in bold refer to a vector notation while underlined and bold symbols refer to matrices.

It should be noted that average referencing is first applied on the EEG samples as a pre-processing step. This is done to  avoid having an excessive weight on a single channel when choosing the optimal inverse solution (more details can be found in section IV-A). 

{\rev 
As mentioned previously, the inverse model is used first to find the number, location and moments of the dipoles that create the electrical activity measured on the scalp. Then the forward model is used to compute, using these dipoles, the measured voltages on certain locations on the scalp. 

In the next paragraphs, the model that maps dipoles to electrical activity is first explained then the steps used to find the optimal model parameters and compute the approximations of the recordings are presented. 
}

\subsubsection{Definition of the Model}

The data vector {\rev of scalp recordings} at time sample $n$, $\mathbf{s}[n]$, is equal to:
$$
 \mathbf{{{s}}}[n] =
\left[ \begin{array}{c}
{s}_{1} [n]\\
{s}_{2} [n]  \\
\vdots\\
\mathbf{s}_{M} [n] \\
\end{array} \right],
$${\rev where the subscript indicates the electrode number.}

{\rev The {\em inverse model}~\cite{bio_book} defines the production of this data vector
 using the following modeling equation:} 
\begin{equation}
\mathbf{s}[n] = \mathbf{\mathcal{K}}[n] {\mathbf{g}}[n] + noise\label{eq:1}
\end{equation}

{\rev This equation links the dipole moments, locations and their number to the recorded data: 
${\mathbf{g}}[n]$ is the {\em dipoles moments vector}, which contains, for each dipole, its moments (in 3 directions); $\mathbf{\mathcal{K}}[n]$, known as the {\em lead field matrix} maps the dipole moments to the voltage they produce at a given electrode position on the scalp; and the number of dipoles is determined by the size of these matrices (for $N_d$ dipoles, ${\mathbf{g}}[n]$ is a length-$3N_d$ vector).
At any given time instant $n$, the model parameters are hence:
\begin{itemize}
  \item the number of dipoles $N_d$;
  \item the dipole moments, concatenated in the dipoles moment vector ${\mathbf{g}}[n]$;
  \item the dipole locations in 3 dimensions; these are embodied in the lead field matrix $\mathbf{\mathcal{K}}[n]$, which also depends on the electrode locations (each row of $\mathbf{\mathcal{K}}[n]$ corresponds to one electrode location) and on the head model used. Once the number of dipoles, their spatial locations and the head model are known,  $\mathbf{\mathcal{K}}[n]$ can be computed, as detailed in 
\cite[ch. 11]{bio_book}.
  \end{itemize}
}

\subsubsection {Modeling the EEG recordings}

{\rev Given a set of measured scalp recordings, the main purpose of the inverse model is to find the combination of model parameters that best reproduce this data through modeling equation~\eqref{eq:1}.}


To {\rev  this end}, we first construct a  discrete set of possible dipole locations. This is referred to as the {\rev \em sampling grid}. {\rev This discretization in space  of the interior of the head model is used in order to reduce the complexity of the fitting process.}
The grid is constructed based on the chosen volume conductor model and the electrode positions. Spatial sampling with uniform spacing in the $x$, $y$ and $z$ directions is applied inside the head volume conductor.

{\rev The inverse model process consists in choosing the combination of dipoles on the sampling grid (number and location) and their respective 3-D moments that best fit the measured data.}

{\rev For a given equation maximum umber of dipoles $N_d$, one can define  the set ${\cal O}$ of all combinations of at most $N_d$ dipole locations. For a given combination of dipoles $c \in {\cal O}$, we can measure the corresponding lead field matrix $\mathbf{\mathcal{K}}_c[n]$.} 



{\rev Neglecting the noise in~\eqref{eq:1}, for a given combination of dipole locations $c \in {\cal O}$ and scalp measurement $\mathbf{s}[n]$, one can estimate the corresponding  moments vector as }
\begin{equation}
{\mathbf{\bar {g}}}_c[n] =  \mathbf{\mathcal{K}}_c^{+}[n]  \mathbf{s}[n],\label{eq:2}
\end{equation}
where  $\mathbf {\mathcal K}_c^{+}[n]$  is the pseudo-inverse of  $\mathbf {\mathcal K}_c[n]$.
The pseudo-inverse can be calculated using the singular value decomposition (SVD) of the lead field matrix $\mathbf {\mathcal K}_c[n]$:
\begin{equation}
\mathbf {\mathcal K}_c[n] =  \mathbf{{\underline U}} \mathbf{{\underline \Sigma}} \mathbf{{\underline V}}^{T}
\end{equation}
 where the columns of $\mathbf{{\underline U}}$ are the eigenvectors of $\mathbf {\mathcal K}_c[n]  \mathbf {\mathcal K}^{T}_c[n] $ and the eigenvalues of $\mathbf {\mathcal K}_c[n] \mathbf {\mathcal K}^{T}_c[n]$  are the squares of the singular values in $\mathbf{{\underline \Sigma}} $. 
 
 The pseudo-inverse is then defined as:
 \begin{equation}
\mathbf {\mathcal K}^{+}_c[n] =  \mathbf{{\underline V}} \mathbf{{\underline \Sigma}^{-1}} \mathbf{{\underline U}}^{T}
\end{equation}

Using the estimate of the moments and the lead field, we can calculate an estimate of the potentials for each combination of dipoles $c$:
\begin{equation}
\mathbf{\bar{s}}_c[n] =  \mathbf{\mathcal{K}}_c[n] {\mathbf{g}}_c[n]		       
\end{equation}

As mentioned previously, our aim is to find the best approximation of the measured voltages, $\mathbf{{s}}[n]$ from  a set of possible potentials, $\mathbf{\bar{s}}_c[n]$, at time sample $n$.
 
 The {\rev squared} difference between the measured and modeled data for a given dipole configuration $c$  is used as criterion {\rev when finding the best fit combination $c$  in the inverse model. It is given by}:
{\rev \begin{eqnarray}  
 d_c[n] &=& (\mathbf{s}[n]-\mathbf{\bar{s}}_c[n])^T(\mathbf{s}[n]-\mathbf{\bar{s}}_c[n])\nonumber\\
 &=& (\mathbf{s}[n]-\mathbf{\mathcal{K}}_{c}[n] \bar{\mathbf{g}}_{c}[n])^T(\mathbf{s}[n]-\mathbf{\mathcal{K}}_{c}[n] {\bar{\mathbf{g}}}_{c}[n])\nonumber\\
 &=& (\mathbf{s}[n]-\mathbf{\mathcal{K}}_{c}[n] \mathbf{\mathcal{K}}_{c}^{+}[n]  \mathbf{s}[n])^T(\mathbf{s}[n]-\mathbf{\mathcal{K}}_{c}[n] \mathbf{\mathcal{K}}_{c}^{+}[n]  \mathbf{s}[n])\nonumber\\
 &=& \mathbf{s}^T[n](\mathbf{I}-\mathbf{\mathcal{K}}_{c}[n] \mathbf{\mathcal{K}}_{c}^{+}[n])^T(\mathbf{I}-\mathbf{\mathcal{K}}_{c}[n] \mathbf{\mathcal{K}}_{c}^{+}[n])\mathbf{s}[n].
\end{eqnarray}}

{\rev Based on the above, the best combination $c^* \in \cal O$ at a given time $n$  is such that
\begin{equation}
  c^* =\arg \min_{c \in \cal O} d_c[n]
  \label{eq:3}  
\end{equation}}


Thus, having simply the electrode positions and an appropriate head model, one can estimate the dipoles' locations then lead field and the dipoles' moments ${\bar{\mathbf{g}}}$ in order to finally compute an approximation of the measured EEG data.


The solution to the inverse problem gives us the optimal dipoles positions and moments that best map to the observed signals. In compression, based on the solution of the inverse problem, the forward model is used to compute the approximations of the signals. Knowing simply the electrode positions, the agreed-upon head model and the dipoles positions and moments, we are able to compute the lead field matrix and thus the estimated potentials, $\mathbf{\bar{s}}[n]$, for each time sample $n$.

\section{Compression using the Forward Model Algorithm}
In compression, one should be able to control the rate of compression and thus the level of distortion added to the data. Using the algorithm shown in the previous section, one can specify the number of dipoles, $N_d$, to use in the fitting process.

In the context of compression, the EEG measurements are modelled as the solution potentials of the forward problem. The parameters that define this model are the head model and the position and moments of the dipoles. Testing has shown that using a standard head model and a relatively low number of dipoles, one can get a good approximation of the EEG signals. In this section we focus on coding techniques used to represent the dipoles' characteristics and the residual error between the actual and estimated potentials.

As seen in the previous section, using simply the electrode positions, the agreed-upon head model and the dipoles' positions and moments, we are able to compute the lead field matrix, $ \mathbf{\mathcal{K}}[n]$. Then, using the values of $\bar {\mathbf{g}}[n]$, we are able to compute $\mathbf{\bar{s}}[n]$. Thus, in compression, the dipoles' positions and moments need to be shared between the encoder and decoder for the computation of the approximated signals  $\mathbf{\bar{s}}[n]$. Thus, in our system, the overhead is composed of the dipoles' positions and moments.

EEG signals are segmented into different matrices. The EEG matrix $\mathbf{\underline{s}}[i]$ is composed of samples ${s}_l[i,n]$, of matrix index $i$, with $l$ going from $1$ to the number of channels $M$ and of length $N$ samples, i.e. $n$ varies between $1$ and $N$. In each matrix, dipoles are considered to be fixed which means that the location $p_o$ of each dipole $o$ is considered to be constant for all values of $n$ within the same matrix at index $i$. Thus, for each EEG block $\mathbf{\underline{s}}[i]$, we define a set of dipole positions $\mathbf{p}[i]$, composed of the positions of each dipole $o$, that do not change within the block $i$.

It should be noted that when applying dipole fitting on EEG blocks of length $N$, moments of the dipoles are generated for all time samples $n$ varying from $1$ to $N$. Thus the dipoles moments are now represented by the matrix $\underline{\mathbf G}[i]$ that corresponds to the dipoles of EEG block $\mathbf{\underline{s}}[i]$. The size of the matrix $\underline{\mathbf G}[i]$ is equal to $3N_{d}$-by-$N$.

\begin{figure}[h!]
\centering        
\includegraphics[scale = 1]{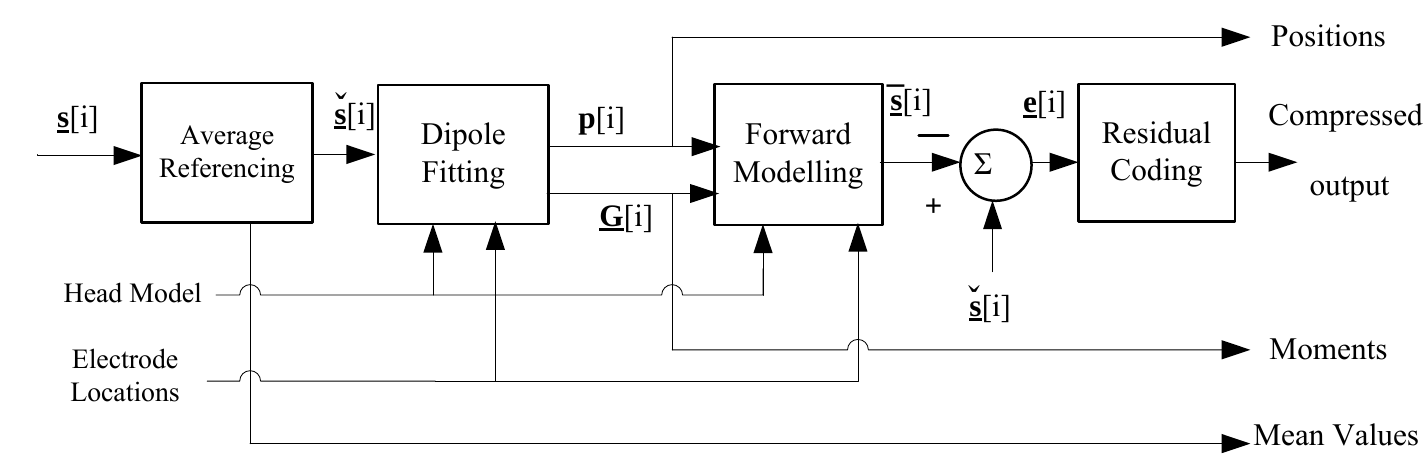}
\caption{Block Diagram of the Overall Compression System.}
\end{figure}

Figure $1$ shows the block diagram of the overall compression system with the EEG matrix $\underline{\mathbf{s}}[i]$ as input. Dipole fitting is first used to compute the position of the dipole, $\mathbf{p}[i]$, and the moments $\underline{\mathbf{G}}[i]$, along the three directions $x$, $y$ and $z$. Then, forward modelling computes the estimated signals from the dipole and moments. This step uses the same head model and electrode locations used in dipole fitting to compute the dipole grid. Afterwards, residual coding is applied on $\underline{\mathbf{e}}[i]$. 

\subsection{Average Referencing}

For each EEG matrix $\mathbf{\underline{s}}[i]$ of index $i$, dipole fitting attempts to model the recordings at each time sample $n$ along all channels $l$. These vectors correspond to the vertical components of the EEG matrix $\mathbf{\underline{s}}[i]$ of length $M$ each.  

Before performing dipole fitting, EEG data is average referenced. This means removing the means of the vectors $\mathbf{s}[i,n]$ where the channel index $l$ varies and $n$ and $i$ are kept constant:
\begin{equation}
\mathbf{\check s}[i,n]= \mathbf{s}[i,n] - \mathbf{\hat \mu}_{s}[i,n]
\end{equation}
where $\mathbf{ \mu}_{s}[i,n] = \frac{\sum_{l=1}^{M} \mathbf{s}_l[i,n]}{M} $, which is equal to the mean of the vector $\mathbf{s}[i,n]$ and $\mathbf{\hat \mu}_{s}[i,n]$ is the quantized version of $\mathbf{\mu}_{s}[i,n]$. This is done in order to have average referencing along all time segments. When performing average referencing, we avoid having an excessive 
weight on an arbitrary single channel (i.e. the reference channel). However this   
adds an overhead of $N \times B_{\mu}$, per block, with $B_{\mu}$ equal to the number of  bits used to quantize the mean values.

It should be noted that quantized values of the mean are used in order to guarantee having the same values at both the coder and decoder's sides. 

\subsection{Coding the Dipole Moments}
The inverse problem of finding the dipoles that best fit the measurements finds as solution the positions and the moments of each dipole. Moments are vectors, in the $x$, $y$ and $z$ directions, of length equal to the number of time samples of the signals we are trying to model. Thus, each dipole is described with a $3$-by-$1$ position vector and a $3$-by-$N$ moments matrix, where $N$ is the chosen segment length of the EEG channels. The compression ratio depends on the number of dipoles used to fit the model, the coding technique of the positions and moments of these dipoles, and the coding technique used for the residual.

\begin{figure}[h!]
\centering        
\includegraphics[scale = 1]{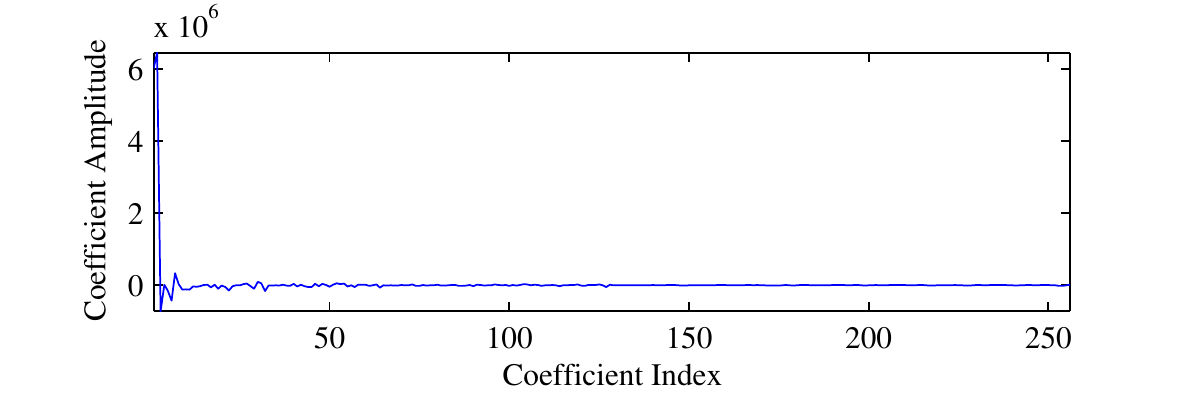}
\caption{Example of the DWT coefficients of a dipole moment with $N=256$.}
\end{figure}

When analysing the characteristics of the moments, we notice that a lot of energy
compaction can be achieved when performing DWT on these moments. This is highlighted in Fig. $2$ when $N=256$ and Daubechies (db1) wavelet is used. Thus, appropriate zero-tree coder such as SPIHT in $1$D \cite{Lu99waveletcompression} \cite{ecg2}  applied on the DWT coefficients yields low values of distortion when using a constant bit rate of $3$ bps. Detailed results are shown in section VI of the article.

\subsection{Coding the Residuals}
Residuals of the dipole modelling part of the compression algorithm are equal to:
\begin{equation}
{e}_l[i,n]= {\check s}_l[i,n] - {\bar s}_l[i,n],
\end{equation}
where ${\check s}_l[i,n]$ is the mean averaged EEG sample of matrix index $i$, at channel $l$ and time sample $n$; and ${\bar s}_l[i,n]$ is the modelled sample also of matrix index $i$, at channel $l$ and time sample $n$ obtained after applying the forward model.

In this section, we present two different methods to code the residual. First, we propose a coding method targeted for Discrete Cosine Transform (DCT) coefficients and based on significance and refinement passes, which is inspired by SPIHT coder applied on DWT coefficients \cite{Lu99waveletcompression} \cite{ecg2}. This method performs well when dipole fitting is able to well model the measured signals. However, for certain types of signals, it is hard to obtain a good approximation. Thus, for such cases, a different residual coding method is also suggested. This method aims at further decorrelating the computed residuals. Details of both methods are explained in this section.

\begin{enumerate}
\item Coding of DCT Coefficients \\
When a good approximation is achieved, residuals have most of the energy concentrated in the low frequency coefficients. 
In the suggested method, an M-point DCT is applied on each residual channel $l$ of $\mathbf{e}_l[i]$ and the resulting coefficients are referred to as $\mathbf{D}_l[i]$. It should be noted that $\mathbf{e}_l[i]$ is composed of the residual samples ${e}_l[i,n]$ of channel $l$ and block index $i$, with $n$ going from $1$ to block size $N$. 

We first combine the DCT coefficients of the residuals from their $2$D arrangement, where the two dimensions are time sample and channel, into a $1$D arrangement, where $D_l[i,n]$ is the DCT coefficient of the residual of channel $l$, matrix index $i$ and time sample $n$: the coefficients of matrix $\underline{\mathbf{D}}[i]$ are combined in a way to have the first coefficient of channel $l = 1$, $D_1[i, 1]$ followed by $D_l[i, 1]$ with $l$ going from $2$ to $M$. Then we add the DCT coefficients at $n = 2$, $D_1[i, 2]$, followed by $D_l[i, 2]$  with $l$ going from $2$ to $M$, and so on... Figure $3$ shows the suggested arrangement. 

The resulting $1$D arrangement of the coefficients of the EEG matrix at index $i$ is denoted as a one dimensional vector $\mathbf{D}_{1\mathrm D}[i]$ composed of coefficients ${D}_{1\mathrm D}[i,m]$ with $m$ between $1$ and $M \times N$. ${D}_{1\mathrm D}[i,m]$ is then coded sequentially as follows.

\begin{figure}
\centering        
\includegraphics[scale=0.9]{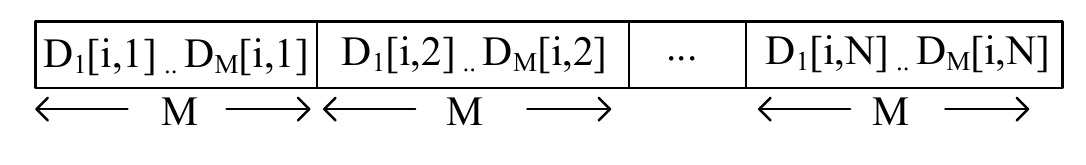}
\caption{Suggested arrangement of the DCT coefficients of all $M$ residuals in $1$D.}
\end{figure}

First we compute the power of $2$ coefficient, $n_p$, of the maximum value of $\mathbf{D}_{1D}[i]$:
\begin{equation}
n_p = \lfloor\log_2(\max(|\mathbf{D}_{1\mathrm D}[i]|)) \rfloor;
\end{equation}
we take the initial threshold to be equal to:
\begin{equation}
T = 2^{n_p};
\end{equation}

we then start looking at each coefficient of  $\mathbf{D}_{1\mathrm D}[i]$ starting from index $1$ to $M\times N$. For each coefficient at index $m$, we test for significance  by comparing to $T$. If the absolute value of ${D}_{1\mathrm D}[i,m]$ is greater or equal to $T$, we output a $1$ bit, followed by the sign bit, $1$ for negative and $0$ for positive values. We then add ${D}_{1\mathrm D}[i,m]$ to the significance list. If the absolute value is smaller than $T$, then we only output a $0$ bit. 

Refinement is also applied directly when checking for significance. At each index $m$, before checking for significance with respect to $T$, we check if ${D}_{1\mathrm D}[i,m]$ is already in the significance list. It if it's not, then we apply the significance test and output the corresponding bits. If it has already been added to the significance list, we check for refinement. We output a $0$ bit if a refinement of $T/2$ needs to be added and a $1$ if a refinement of $-T/2$ should be added.  

After having coded all coefficients in $\mathbf{D}_{1\mathrm D}[i]$, we divide $T$ by $2$ and restart checking for significance and refinement for each index up to $M \times N$. The coder needs to send the length of $\mathbf{D}_{1\mathrm D}[i]$ ($M$ and $N$) and the initial value of $n_p$ to the decoder.

Table $1$ shows an example of applying this method on coding the following values of $\mathbf{D}_{1\mathrm D}[i]$: 
$\mathbf{D}_{1\mathrm D}[i]$ =  [$36$  $-18$   $-8$   $4$   $10$   $-5 $  $1$].

The bitstream generated by the suggested coding results in large sections of zeros indicating non-significance. For this reason, Run-Length-Coding (RLC) is applied and the count of zeros and ones are coded using adaptive arithmetic coding.

\begin{table}
 \caption{Examples of output of significance/refinement algorithm for DCT coefficients\\ $\mathbf{D}_{1\mathrm D}[i]$ =  [$36$  $-18$   $-8$   $4$   $10$   $-5 $  $1$].}   
 \begin{center}
\begin{tabular}{ c c c }

 $T$ & Output stream & Decoded stream \\
 \hline \hline
  $32$ & 100000000 & [ $32$  $0$ $0$  $0$ $0$ $0$ $0$]\\
  $16$ & 011000000 & [$40$ $-16$ $0$ $0$ $0$ $0$ $0$ $0$] \\
  $8$ & 111101000 & [$36$ $-20$ $-8$ $0$ 8 $0$ $0$] \\
 $4$ & 000100110 & [$38$ $-18$ $-6$ $4$ $10$ $-4$ $0$] \\
  \hline  

\end{tabular}
\end{center}
\end{table}

\item ARX Modeling of the Residual \\
For certain segments of recording, dipole fitting is not able to extract the correlation between the different channels and the calculated signals are highly distorted compared to the measured signals. This occurs mostly in recordings where sampling frequency is low and there is a lot of activity in the high frequency band. In such cases, it becomes hard for dipole fitting to model the recorded signals well.

In such cases, we have devised another coding method based on autoregressive modelling  with exogenous inputs (ARX) and single output of the residuals from the inverse/forward modelling to predict the different channels of these residuals. This method explores the redundancy still present after dipole fitting and tries to model the channels using other channels in the residuals.

ARX prediction is a way to de-correlate the channels using a small number of  channels. It is a prediction model with multiple inputs and single output. To build this model, the first question is to choose which channel to predict from a given set of channels. One possibility would be to use as predictors channels that are physically close to the channel to be predicted, using Hjorth graph \cite{1180083}; our initial tests revealed that, because of the nature of the signals at hand, this method does not give good performance. We instead use clustering in order to appropriately choose the different inputs and outputs of the ARX blocks. Clustering allows us to group channels that are mostly correlated together. Thus, we choose the inputs and outputs of the ARX coder from the same cluster in order to guarantee certain correlation that will be exploited in ARX modelling. 

The suggested method first performs clustering of channels based on the L-$2$ norm distances between the channels. In this clustering method, centroids are first chosen randomly from the channels and initial clusters are formed by associating each channel to the least distant centroid. The optimal centroid for each cluster is then chosen as the member of the cluster having the minimum sum of distances with all other members of the same cluster. 

After having grouped all $M$ channels into different clusters, we find a second centroid per cluster by taking the channel in each cluster that is the least distant from all other members of the same cluster. Thus, each cluster has two different centroids. We keep repeating the process until we have $N_c$ centroids for each cluster. The value of $N_d$ depends on the chosen number of clusters and total number of channels.
 
The centroids are taken as the inputs to the ARX system to model each member of the corresponding cluster. In this method $L[l, j]$ refers to the centroid indices that correspond to channel at index $l$. Thus the channel at index $l$ is modelled using the channels at indices $L[l, j]$ with $j$ from $1$ to $N_c$, the chosen number of centroids per cluster. Centroids are first coded using the DCT-based method suggested in section IV-C-1. The decoded centroids are used as inputs in the ARX blocks. This guarantees that both the encoder and decoder use the same inputs in the ARX blocks.

The equation of the predicted output of the ARX block for EEG residual $e_{l}[i,n]$ of channel $l$, block $i$ and sample $n$ using $N_{c}$ DCT-coded residuals $\hat e_{L[l,1]}[i,n]$, $\hat e_{L[l,2]}[i,n]$, ..., and $\hat e_{L[l,N_{c}]}[i,n]$, with filter length $B$, is the following:
 
\begin{eqnarray}
\hat{e}_{l}[i,n] &=& w_{1,B} \hat{e}_{L[l,1]}[i,n-B+1] +w_{1,B-1} \hat{e}_{L[l,1]}[i,n-B+2] +   \hdots +  w_{1,1} \hat{e}_{L[l,1]}[i,n] +  \nonumber \\
 & &  w_{2,B} \hat{e}_{L[l,2]}[i,n-B+1] +w_{2,B-1} \hat{e}_{L[l,2]}[i,n-B+2] +   \hdots +  w_{2,1} \hat{e}_{L[l,2]}[i,n] + \nonumber\\
 & &  ... +   w_{N_c,B} \hat{e}_{L[l,N_c]}[i,n-B+1] + \hdots +  w_{N_c,1} \hat{e}_{L[l,N_c]}[i,n] 
\end{eqnarray}

The filter coefficients are the following:
 
\begin{equation}
\mathbf w[i] = \left[
\begin {array}{cccccccccc}
w_{1,B} & \hdots &  w_{1,1} & w_{2,B}  & \hdots &  w_{2,1}   & \hdots & w_{N_c,B}   & \hdots & w_{N_c,1} 
\end {array}
\right]
\end{equation}

The filter uses  $B$ samples of each of the  $N_c$ input channels for each of the channel to be decorrelated. The overall number of coefficients is equal to the number of channels used in the decorrelation multiplied by the value of $B$. These coefficients are quantized and sent to the decoder. A value of $B=1$ is chosen in this method since it gives good prediction while ensuring a small overhead. In addition, using a  small number of filter coefficients limits the effect of quantization errors.  

The filter coefficients are calculated using the pseudo-inverse method. The input matrix, $\mathbf{{\underline X}}[i]$, of the filter is formulated using the $N_c$ centroids. When $B$ is set to a value of $1$, this matrix is equal to:
\normalsize{
\begin{equation}
\mathbf{{\underline X}}[i]  = \left[ \begin{array}{ccccccccccccc}
 e_{L[l,1]}[i,1] & e_{L[l,2]}[i,1]  &  \hdots  &  e_{L[l,N_{c}]}[i,1]  \\
  e_{L[l,1]}[i,2] & e_{L[l,2]}[i,2]  &  \hdots  &  e_{L[l,N_{c}]}[i,2]  \\
\vdots   & \vdots  & \vdots    & \vdots    \\
 e_{L[l,1]}[i,N]   & e_{L[l,2]}[i,N] & \hdots  & e_{L[l,N_{c}]}[i,N]  \\
\end{array}\right]
 \end{equation}}\\
 
The desired ARX filter output is equal to:
\begin{equation}
\mathbf{{e}}_{l}[i]=  \left[ \begin{array}{c} e_{l}[i,1] \\ e_{l}[i,2] \\ e_{l}[i,3] \\ \vdots \\ e_{l}[i,N] \end{array} \right]
\end{equation}\\
The filter coefficients are found using the desired output and the pseudo-inverse of the input matrix $ \mathbf{{\underline X}}^{+}$:
\begin{equation}
\mathbf w^{T}[i] = \mathbf{{\underline X}}^{+}[i]  \mathbf{e}_{l}[i]
\end{equation}

Pseudo-inverse, in addition to guaranteeing to find the minimum (Euclidean) norm solution to a system of linear equations, is also relatively fast and simple. The pseudo-inverse gives the solution with the smallest sum of squared filter weights which is desirable since these weights should be quantized and sent.

The predicted output is equal to:
\begin{equation}
{\mathbf{\bar{e}}}_{l}[i]= \mathbf{{ X}}[i]  \mathbf {w}^{T}[i]
\end{equation}\\
And the ARX prediction residual is:
\begin{eqnarray}
\mathbf{y}_{l}[i] &=& \mathbf {e}_{l}[i] - \mathbf{\bar{e}}_{l}[i] 
\end{eqnarray}

\begin{figure}[h!]
\centering        
\includegraphics[scale = 1.1]{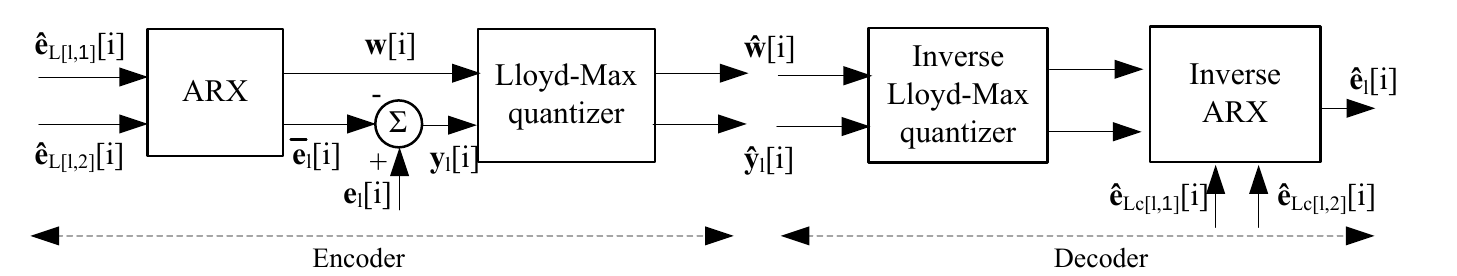}
\caption{Block Diagram of Coding EEG segment $\mathbf e_{l}[i]$ at block $i$ of channel $l$ using the  ARX Modelling method.}
\end{figure}

Figure $4$ shows the block diagram for coding the EEG residual at channel index $l$ with $N_c =2$. $\mathbf{ \hat{e}}_l[i]$ refers to the decoded residual at channel $l$. Lloyd-Max quantizers are used to quantize the residuals and the filter coefficients. The number of bits used to quantize the filter coefficients is kept constant whereas for the ARX prediction residuals, this number varies and is used to control the compression rate.
\end{enumerate}

\subsection{Smoothness Measure and Conditional Coding}
As explained above, depending on how well dipole fitting is able to model the measured data, different coding techniques should be used to code the residuals. When dipole fitting  models the data well, it is able to extract the redundancy between the different channels, and thus the method based on ARX prediction incurs too much overhead for the gain it brings for coding the residuals. For such cases, the DCT-based method is able code the residuals with low distortion at very low bit rates. 

However, as previously mentioned, for matrices with a lot of high frequency content, dipole fitting is not able to model the signals well.
In such cases, the residuals are not well de-correlated and the resulting matrix is not smooth. By smooth we mean that the adjacent samples, both vertically and horizontally, vary a lot. These types of matrices have high energy in the high frequency band. For such matrices, ARX modelling is able to further de-correlate the channels and provide better coding. For this reason, choosing the appropriate coding technique is based on a certain smoothness measure calculated on the residual matrix. We choose the following smoothness measure for the residual matrix $\mathbf e_l[i]$ at index $i$ for all channels with $l$ from $1$ to $M$:
\begin{eqnarray}
\rho[i] &=& \frac{1}{M} \sum_{l=1}^{M}[ \frac{\sum_{n=1}^{N/4} |D_l[i,n]|}{\sum_{n=N/4+1}^{N} |D_l[i,n]|}]
\end{eqnarray}

In this equation $N$ is the block size and $M$ is the number of channels used in the coding. This smoothness measure is calculated from the DCT coefficients of $\mathbf e_{l}[i]$,  $\mathbf D_{l}[i]$.
Based on values of $\rho[i]$ we choose a certain threshold $T_{\rho}$ that determines whether the residual matrix is considered smooth or not. When $\rho[i]$ is greater or equal to $T_{\rho}$, the matrix can be considered smooth and the DCT-based coder is used, otherwise, the ARX-based coder is used.

\section{Datasets}
This section presents a detailed description of the datasets used in testing the compression performance
of the suggested method.
\subsection{Dataset $1$- MIT dB}
The first set of data, CHB-MIT Scalp EEG Database, was collected at the Children's Hospital Boston. This dataset consists of EEG Scalp recordings from pediatric subjects with intractable seizures \cite{physiobank}. Testing was done on $11$ patients\footnote{Patients $01$ to $05$, $07$, $09$, $11$, $18$, $20$, $22$ and $23$, were used in the testing.}, with ages varying between $6$ and $22$ years. Each patient can have multiple seizures { during recording}, ranging from a single seizure to $7$ seizures. This dataset is sampled at $256$ Hz. Recording is done over $25$ channels with bipolar montage and $16$ bits are used in the recording's precision.

\subsection{Dataset $2$- MNI dB}
To test the compression performance of the system, testing is also done on recordings acquired at the Montreal Neurological Institute (MNI dB) from $9$ patients using $29$ channels, $200$ Hz sampling frequency and $16$ bits were also used in the recording's precision. It should be noted that these recordings were acquired for the purpose of this study and are not publicly available. 

\subsection{Dataset $3$- Delorme dB}
The data of the third database is published on-line and recorded by Alain Delorme \cite{data} with $31$ channels, sampled at $1000$ Hz. Patients were shown certain images during the recording and they were required to indicate whether or not the image is familiar. 
The first three records of each of the $8$ patients available in this database are used in the testing.

\section{Results}
\begin{figure}[h!]
\centering        
\includegraphics[scale = 1]{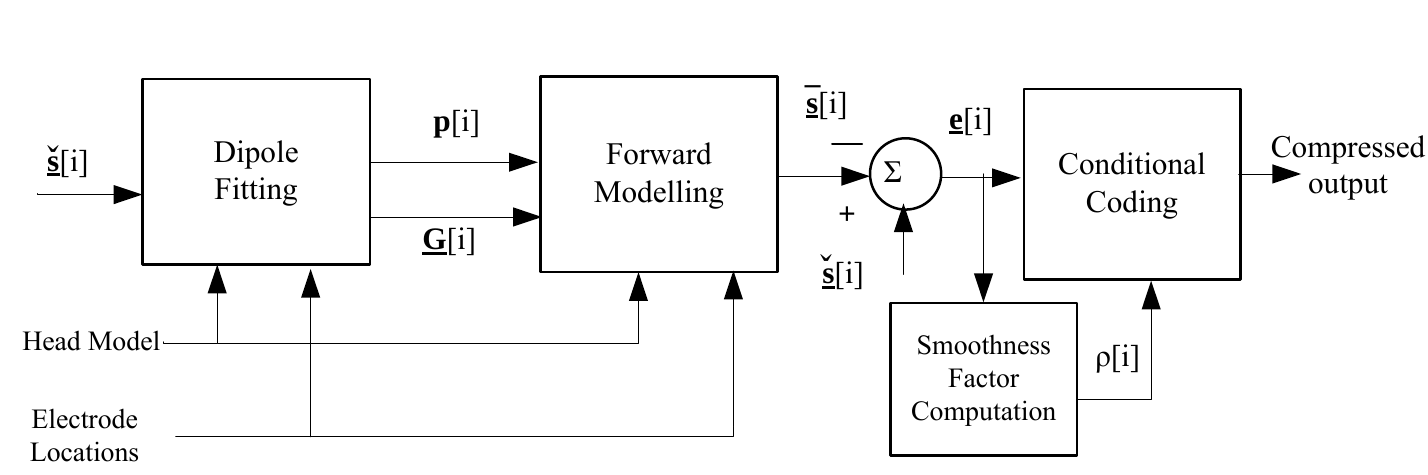}
\caption{Block Diagram of the Overall Compression System.}
\end{figure}

In this section, compression performance of the suggested system tested on three different datasets is presented.

To determine the appropriate head model assumption in the inverse and forward models, the DIPFIT validation study in \cite{eeg_online} is tested on patient $1$ of Delorme dB to compare the effect of the two head
models: concentric spheres with 4 compartments and standard BEM head model. Results show that fitting using the two models results in a small difference in dipole positions. However, there are outliers that account for less than $10\%$ of all dipoles, and are mostly due to bad convergence when using either of the two models \cite{eeg_online}. However, the sBEM model is chosen in our study because it gives slightly better approximation when dipole fitting is not able to converge to an appropriate solution.

Another important parameter to consider in the dipole fitting model is the sampling grid. As explained in section IV, dipole fitting tries to find the best dipole, or combination of dipoles, in the sampling grid, that best maps to the recorded signals. The finer the grid, the more possible dipole locations in the system. To examine the effect of granularity of the grid on the system, we modify the spacing in the $x$, $y$ and $z$ directions used to build the grid. This changes the number of dipoles used in the fitting scenario. The grid scale was varied to include $116$, $181$ (default dipole locations used in EEGLAB), $523$ and $887$ different dipole locations used in the fitting process. There was no substantial difference in distortion for all databases. It should be noted that the finer the grid, i.e. number of dipoles equal to $887$, the larger the delay caused by searching a very large number of possible locations.

Fitting is done using a low number of dipoles in order to obtain an output that is compressed compared to the input. Since we focus only on a low number of dipoles, the complexity of the system is linear in terms of the size of the grid. Thus there is negligible difference in complexity and delay between a grid size of $116$ and $181$. 
For these reasons, we set the number of dipoles to the default value of $181$. However, if a large number of dipoles is used, a smaller grid size of $116$ locations is recommended to reduce the complexity and delay of the system that exponentially increase with the number of dipoles.

Figure $5$ shows the block diagram of the overall compression system with the EEG matrix $\underline{\mathbf{\check s}}[i]$ as input. As previously explained in section IV-A, the matrix $\underline{\mathbf{\check s}}[i]$ is obtained after applying average referencing on the channels of $\underline{\mathbf{s}}[i]$. As mentioned in section IV, dipole fitting is first used to compute the position of the dipole, $\mathbf{p}[i]$, and the moments $\underline{\mathbf{G}}[i]$. Then, forward modelling computes the estimated signals from the dipole and moments. The residual $\underline{\mathbf{e}}[i]$ is found and coded based on the value of $\rho[i]$. The output of the residual coder, along with the position and coded moments are sent to the decoder. 

At the decoder's side, the estimated potentials, $\underline{\mathbf{\bar s}}[i]$ are first computed using the dipole's moments and positions and of course the agreed-upon head model and electrode positions. Afterwards, the decoded residuals are added to these estimated potentials. In order to reconstruct the original signals, $\mathbf{\hat \mu}_s[i]$ should also be added to compensate for the average referencing block. The decoded samples of channel $l$, matrix index $i$ and time sample $n$ are  $\hat{s}_{l}[i,n]$.

The performance parameter used to analyse the results is the percent-root mean square difference (PRD):
 \vspace{-4pt}

\begin{equation}
PRD_{l}[i](\%) = \sqrt{\frac{\sum_{n=1}^{N}(s_{l}[i,n] - \hat{s}_{l}[i,n])^2}{\sum_{n=1}^{N}s_{l}[i,n]^2}} \times 100 
\label{eq:20}
\end{equation}
where ${s}_{l}[i,n] $ is the EEG sample $n$ of segment at index $i$ and channel at index $l$, and $\mathbf{\hat{s}}_{l}[i] $ is the reconstructed EEG segment after compression. Thus $PRD$ is calculated for each EEG channel, $l$, and segment $i$, then the mean over all segments and channels is calculated to reflect the $PRD$ at a certain $CR$.


\begin{figure}[h!]
\centering        
\includegraphics[scale = 0.9]{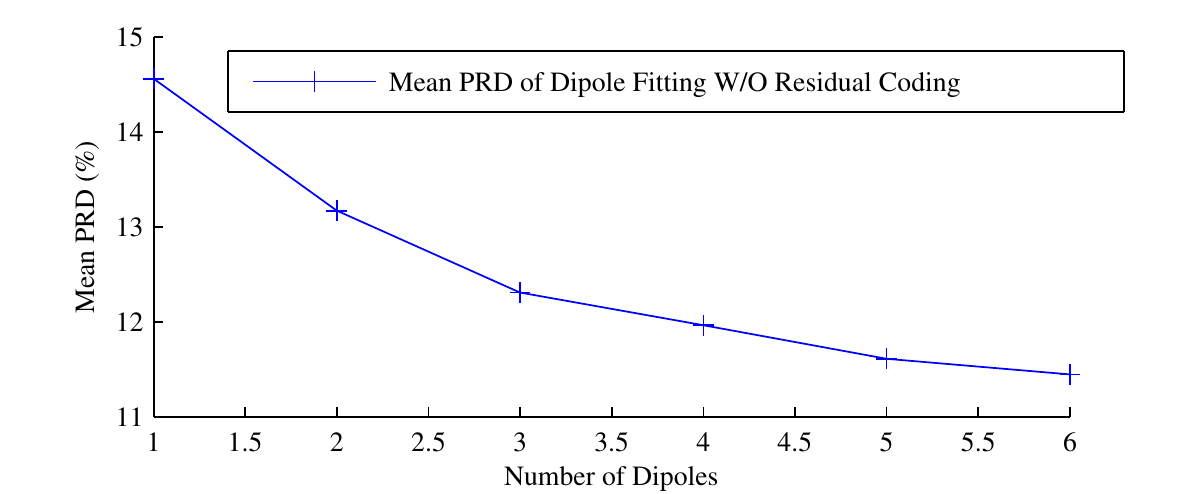}
\caption{Effect of increasing the number of dipoles on the $PRD$ values of the output of the forward model.}
\end{figure}

As previously mentioned, the lower the number of dipoles, the lower the number of moments that need to be sent to the decoder. Testing has shown that the higher the number of dipoles used in the fitting, the more accurate the model 
.  However, when we increase the number of dipoles by $1$, we are actually multiplying the overhead by $2$ since the position vector and the moments matrix double in size. Thus the overhead increases a lot while the improvement in accuracy is small. Thus a low number of dipoles is recommended in order to ensure high compression rates (CRs). In our compression algorithm, we limit the number of dipoles to $1$. 

Figure $6$ shows the effect of varying the number of dipoles on the dipole fitting process tested on patient $1$ of Delorme dB. To compute the $PRD$ values in this plot, we replace $\hat{s}_{l}[i,n]$ by the modelled signal $\bar{s}_{l}[i,n]$ in equation~\eqref{eq:20}. We can directly see from these results that when we increase the number of dipoles, distortion decreases. However, as mentioned previously, the slight improvement caused by using one additional dipole doubles  the overhead and thus increases significantly the bit rate. For this reason, only one dipole is used in our suggested system.

\begin{figure}[h!]
\centering        
\includegraphics[scale = 0.9]{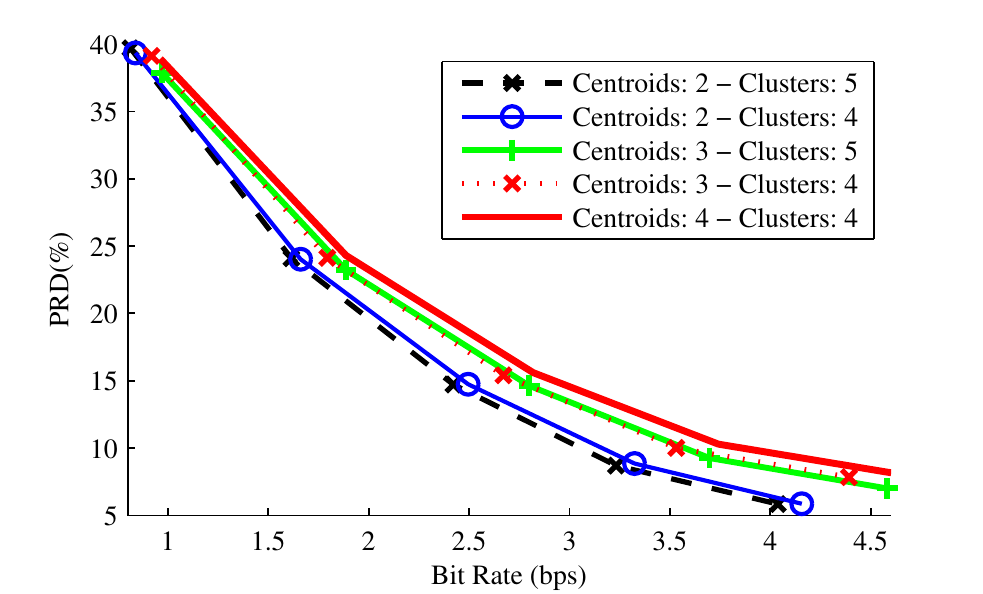}
\caption{Performance Comparison of the ARX-based residual coding method with varying number of centroids and clusters applied on the residuals of Patient 1 of MNI dB.}
\end{figure}

The residual coding method based on ARX and clustering is highly influenced by the number of clusters and centroids used. Fig. $7$ shows the effect of changing these two parameters on the recording of patient $1$ of MNI dB in terms of compression distortion ($PRD$) versus bit rate ($BR$). It should be noted that $4$ bits are used to quantize the filter coefficients and $B$ is chosen to be equal to $1$. Testing showed that increasing the number of previous samples used in prediction does not improve the performance of the system, and introduces a slightly larger overhead.

We can directly see from Fig. $7$ that when increasing the number of centroids, there is degradation in performance. In fact when we increase the number of centroids, more channels are coded using the DCT-based coder. In addition, a higher number of centroids means that inputs that might not be correlated to a certain channel are used in the ARX prediction of that channel. Further increasing both parameters implies that most of the channels are coded using the DCT based coding method and the performance deteriorates at high bit rates. It should be noted that the ARX-based method used to compute the results in the rest of this article uses $5$ clusters and $2$ centroids.

\begin{figure}[h]
\centering
\begin{minipage}{1\textwidth}

\begin{subfigure}[a]{1\textwidth}
\center
\includegraphics[scale = 1]{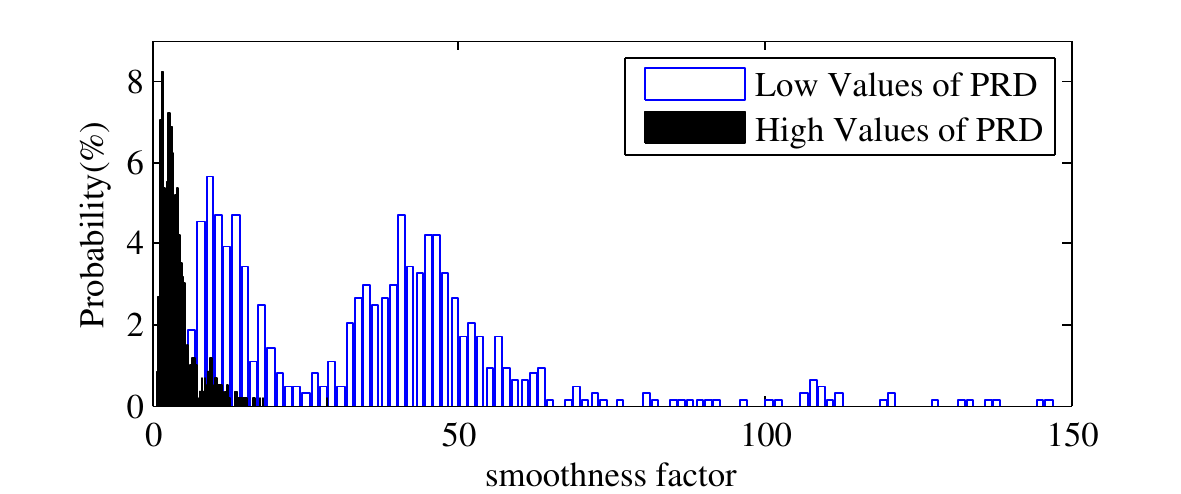}
\caption{Values for Patient $1$ of all three databases.}
 \end{subfigure}  
\end{minipage}
\begin{minipage}{1\textwidth}
\begin{subfigure}[b]{1\textwidth}
\center
  \includegraphics[scale = 1]{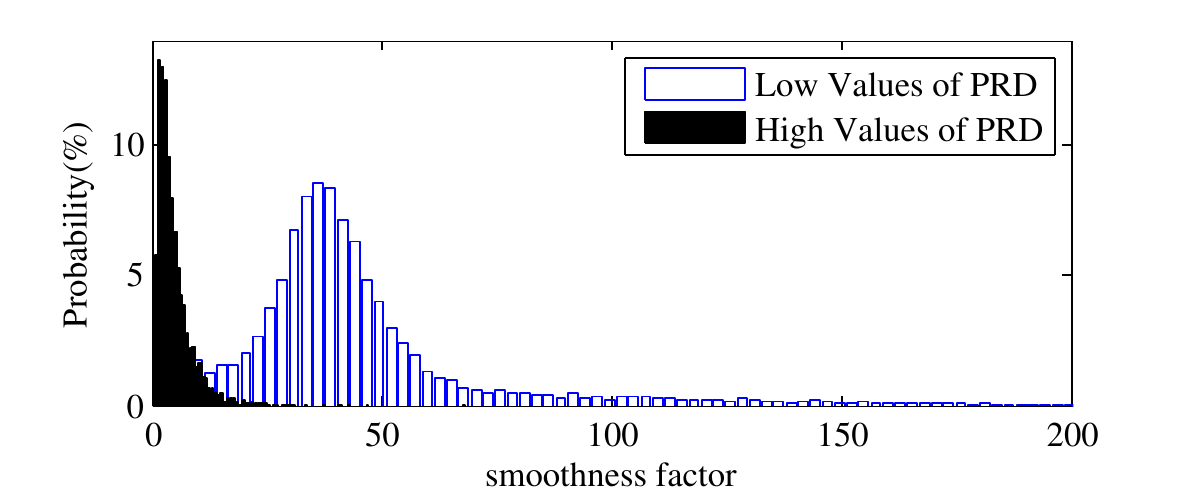}
\caption{Values for all patients of all three databases.}
   \end{subfigure}  
\end{minipage}
 \caption{Distribution of smoothness factor values of the three datasets based on low and high values of $PRD$ for the DCT-based suggested coder.}
\end{figure}

The threshold of the smoothness factor $T_{\rho}$ is calculated and chosen based on the recordings of patient $1$ of all three databases. Depending on the value of $\rho[i]$, the appropriate coding method of the residuals is chosen. When the value of $\rho[i]$ is greater than or equal to the threshold  $T_{\rho}$, residuals are coded using the DCT-based coder, otherwise, they are coded using the ARX-based coder. 

The distributions of the smoothness factor are shown in Figures $8$-(a) and $8$-(b). Figure $8$-(a) shows the distributions when testing is done on only patient $1$ of all three databases, whereas figure $8$-(b) shows the distributions when testing is done on all patients of all three databases. In these figures, the first black plot corresponds to values of $\rho[i]$ where the corresponding $PRD$ is high, whereas the second distribution plot corresponds to  low values of $PRD$. The $PRD$ values are taken for the DCT-based suggested coder for a bit rate of around $1$ and $PRD$ values are compared to a value of $15\%$ to determine if the value is low or high.  

These figures shows that, in fact, for low values of $\rho$, the DCT-based coder does not perform well since there is not much energy in low frequency components. It should be noted that  in figure $8$-(a) the distribution that corresponds to low values of $PRD$ exhibits two separate bell shapes, each corresponding to a certain dataset. Residuals of Delorme dB exhibit very high energy in the low frequencies, therefore the second bell shape corresponds to this dataset while the first bell shape corresponds to the recordings of MIT dB. When taking a larger sample size, i.e. as in Fig. $8$-(b), these two bell shapes merge into one. Both figures have a consistent threshold value below which we obtain high values of distortion.

\begin{figure}[h!]
\centering        
\includegraphics[scale = 0.9]{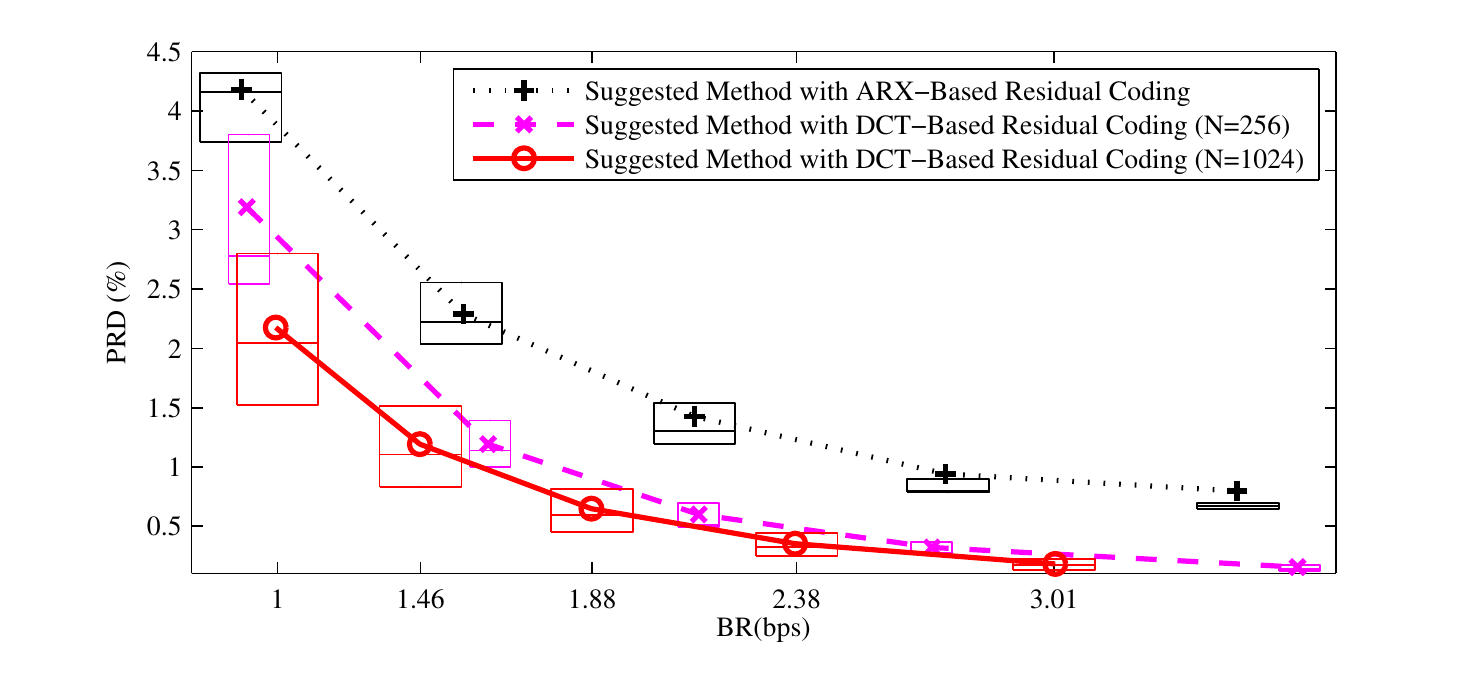}
\caption{Performance Comparison between the different coding methods of the suggested compression system tested on Delorme dB.}
\end{figure}

\begin{figure}[h!]
\centering        
\includegraphics[scale = 0.9]{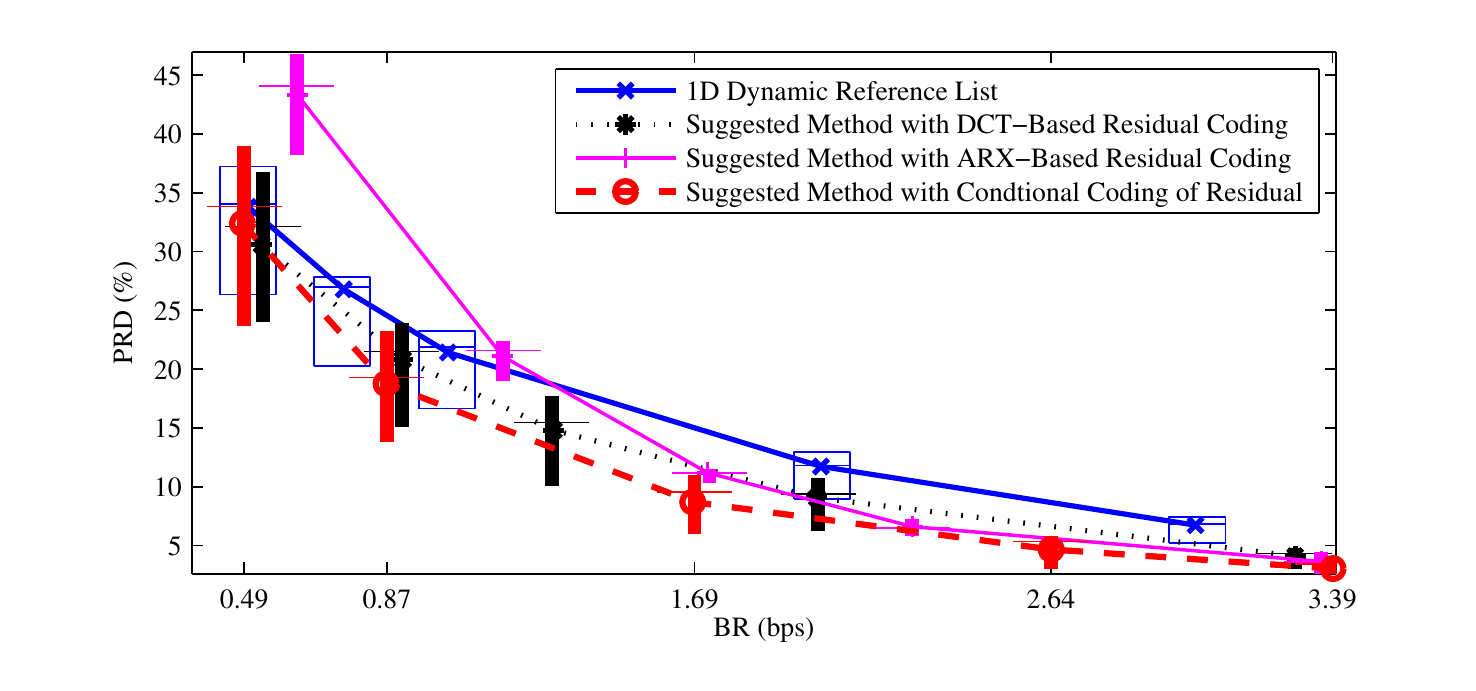}
\caption{Performance Comparison between the different coding methods of the suggested compression system and the previously suggested referential $1$D algorithm \cite{journal} tested on MIT dB.}
\end{figure}

\begin{figure}[h!]
\centering        
\includegraphics[scale = 0.9]{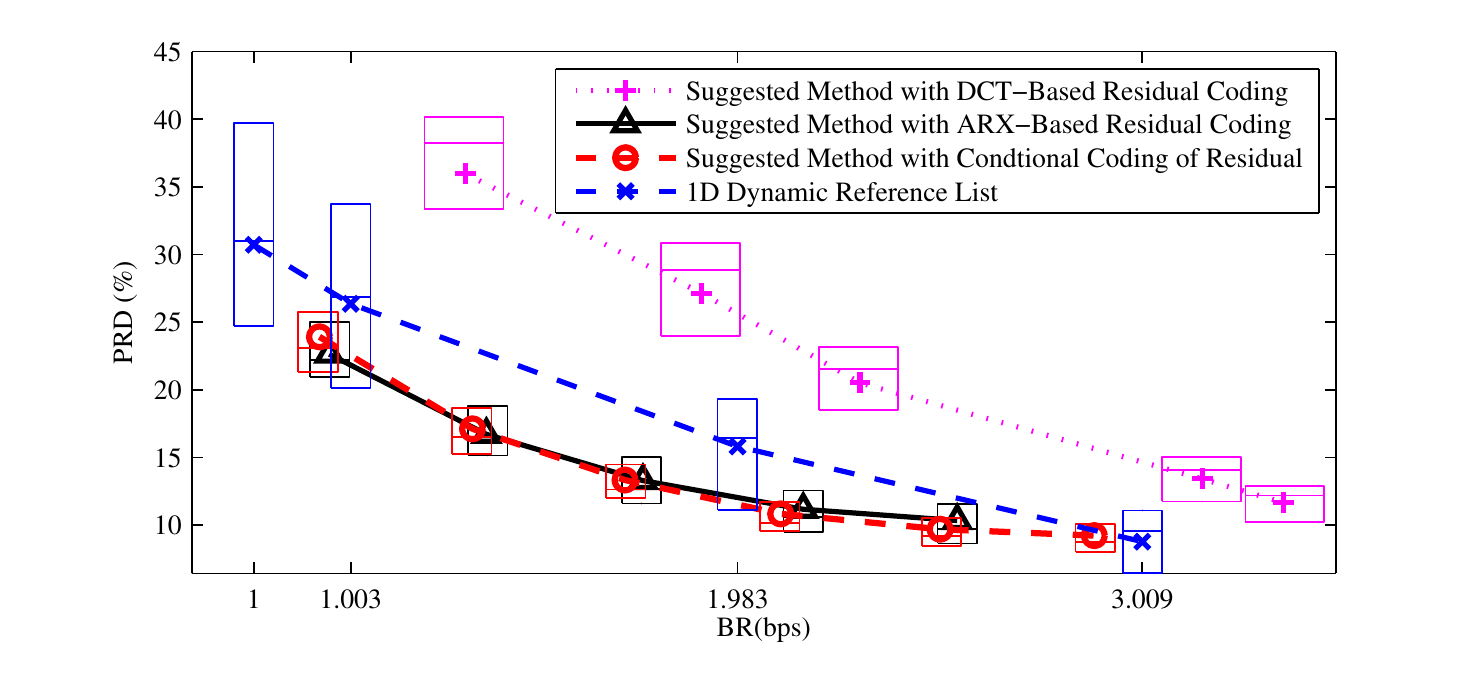}
\caption{Performance Comparison between the different coding methods of the suggested compression system and the previously suggested referential $1$D algorithm \cite{journal} tested on MNI dB.}
\end{figure}

We set the threshold $T_{\rho}$ to be equal to $10$ and run the conditional compression algorithm on all patients  of the $3$ databases. Results of all compression methods and the conditional coding are shown in Figures $9$ to $11$. In these figures, the variations of results between the different patients of each dataset is highlighted using the box-plots of $75$ and $25$ percentiles.

\section{Discussion}

Figure $9$ shows the compression performance of the suggested system when tested on the recordings of Delorme dB. We can directly see from the results that distortion does not vary a lot between the different patients and we can achieve very low $PRD$ for very low bit rates. Both residual coding methods give
good results, however, lower distortion can be achieved with the DCT-based coder with $N=1024$. It should be noted that when comparing with $T_{\rho}=10$, all values of $\rho[i]$ are greater than $T_{\rho}$ and thus conditional coding always chooses the DCT-based coder.

Figures $10$ and $11$ show the compression performance of the suggested system compared to the compression method in \cite{journal}. This method is based on the use of DWT and dynamic reference lists to compute and send the decorrelated sub-band coefficients.

Results of MIT dB are shown in Fig. $10$. Segments of length $N = 1024$ are used in the testing. This value gave better results compared to smaller values such as $256$ and $512$. As previously mentioned, bipolar montage of adjacent electrodes is used in the recording. In order to compute the electrode positions used in dipole fitting, the mid point between the two electrode positions of each bipolar channel is used as approximation. Using these position points, dipole fitting was able to provide a good approximation of the signals.  

These results show that the suggested method gives better performance than the method in \cite{journal}. In addition, results show that since the smoothness factor is relatively high, the dipole fitting residuals are mostly coded using the DCT-based method. However when applying conditional coding of the residuals with $T_{\rho}=10$, we obtain an improvement in performance. 

Results of MNI dB are shown in Fig. $11$. In these plots, $N$ is chosen to be equal to $256$ since it gave better results than higher values. The suggested method with ARX-based residual coding gives the best compression performance. Conditional coding achieves almost the same results as the ARX-based coder. This shows that most segments have low values of $\rho[i]$ and thus dipole fitting did not converge to a good approximation. 

Results show that the method is able to adapt to changes in the characteristics of the data. When recordings are done at high sampling frequencies and for certain evoked potentials, as for Delorme dB, dipole fitting approximates very well the data and the DCT-based coder provides good compression of the residuals.  Near lossless compression can be achieved for these types of recordings at bit rates greater or equal to $2.5 bps$. This corresponds to compression rates lower or equal to $6.4$. 

As mentioned previously, for certain segments of recordings, dipole fitting is not able to generate good approximations of the recorded signals. For such cases, the ARX-based coder of the residuals is able to compensate for this high distortion in the approximated signals. This is highlighted in Figures $10$ and $11$. In Figure $10$, there is a slight improvement when using conditional coding compared to the DCT-based coder. This shows that some of the segments were coded with ARX-based coder which caused an improvement in performance. However, in Figure $11$, almost all the segments of recordings are coded using the ARX-based coder. For MNI dB, most of the segments do not converge to a good approximation using dipole fitting, and thus there is still a lot of redundancy present in the residuals that is taken care of using the ARX-based coder.

\section{Conclusion}
The suggested method is able to provide good compression performance when tested on three datasets of different characteristics. Low distortion values are obtained for bit rates lower than $3 bps$. In addition, near-lossless compression is achieved for a certain type of recordings: event-related potentials. 

Further analysis is needed to study the distortion added to the signals. The suggested compression scheme should still preserve important diagnosis information. It would be important to test abnormality detection systems, like epileptic seizure detection, on both the original data and the compressed output to further analyse the performance of the compression algorithm \cite{Hoda1, mypaper, journal}.

\bibliographystyle{IEEEtran}
\bibliography{EEGBooks}

\end{document}